\documentclass[12pt]{iopart} 
\usepackage{iopams}
\usepackage{amsthm}
\usepackage[utf8]{inputenc}
\usepackage[T1]{fontenc}
\usepackage{float,graphicx,bbm,bm,color}

\begin{document}
\def\newblock{\ }%
\newcommand{\ii}{\mathbbm{i}}
\newcommand{\Z}{\mathbb{Z}}
\newcommand{\R}{\mathbb{R}}
\newcommand{\C}{\mathbb{C}}
\def\Aa{{A{\scriptstyle 1}}}
\def\Ab{{A{\scriptstyle 2}}}
\def\As{{A{\scriptstyle 3}}}
\def\sAa{{A{\scriptscriptstyle 1}}}
\def\sAb{{A{\scriptscriptstyle 2}}}
\def\sAs{{A{\scriptscriptstyle 3}}}
\def\At{\widetilde{A}}
\def\da{{\upsilon_1}}
\def\db{{\upsilon_2}}
\def\du{{\delta_1}}
\def\dt{{\delta_3}}
\def\ga{{\gamma_1}}
\def\gb{{\gamma_2}}
\def\ea{{\eta_1}}
\def\eb{{\eta_2}}
\def\ei{{\eta_i}}
\def\ca{{\chi_1}}
\def\cb{{\chi_2}}
\def\Ec{\mathcal{E}}
\def\Hg{H_{\rm gH}}
\def\sr{s_{\rm r}}
\def\Pim{\Pi_{\rm odd}}
\def\Pip{\Pi_{\rm even}}
\def\sqtrois{{\scriptscriptstyle\sqrt3}}
\def\skxy{{\scriptstyle(k_x,k_y)}}

\title{Bloch classification surface for three-band systems}
\author{G. Abramovici}
\address{Université Paris-Saclay, CNRS, Laboratoire de Physique des Solides,
91405, Orsay, France}
\ead{abramovici@lps.u-psud.fr}

\begin{abstract}
Topologically protected states can be found in physical systems, that show
singularities in some energy contour diagram. These singularities can be
characterized by winding numbers, defined on a classification surface, which
maps physical state parameters. We have found a classification surface, which
applies for three-band hamiltonian systems in the same way than standard
Bloch surface does for two-band ones. This generalized Bloch surface is
universal in the sense that it classifies a very large class of three-band
systems, which we have exhaustively studied, finding specific classification
surfaces, applying for each one.

\end{abstract}

\section{Introduction} 

In recent years, physicists have investigated new quantum states, such as
zero-energy states like Majorana fermions\cite{Read,Kitaev,FuKaneMele},
zero-mass particles associated to Dirac contact
points\cite{Bena,FuchsEpjB,Lih2015} or anyons\cite{Read90}. These quantum states
are remarkable because they are \textit{protected} by \textit{topological}
singularities.

Initiated by theoretical predictions\cite{Wilczek,Haldane,Yang,GirvinMcD,Zhang,
Read,Ezawa,Tanaka,Regnault}, the quest of such topological states has spread
into a larger and larger community of experimentalists and has provided more and
more valid candidates\cite{Hasan,Liuetal,Nayak,Bartolomei,Manfra,HQTBY}.

In most cases, these states are characterized by a quantum integer associated
to some physical flux in real or reciprocal space. Following Gauss-Bonnet
theorem\cite{Allendoerfer,Fenchel,Chern}, this quantum integer is also related
to a path integral around a singularity. The choice of the integrand depends on
which symmetry is relevant in each specific situation.

This approach proves to be very general: the classification of many topological
states can be performed through that of closed paths, using the fundamental
(also called first homotopy) group $\pi_1(\Ec)$, which addresses winding numbers
associated to specific symmetries of the system. Other topological systems need
the second homotopy group $\pi_2(\Ec)$, for which our results are not relevant.
For the classification through $\pi_1(\Ec)$ to be valid, it must be determined
in an abstract space $\Ec$, where all states are represented faithfully.

In primitive theories\cite{Haldane,Regnault,Kalugin}, topological states are
protected by energy gaps. However, more sophisticated cases may happen%
\cite{Tanaka,Abramovici} in three-band systems, where one gap closes. In such
situations, using $\pi_1(\Ec)$ to determinate winding numbers proves very
efficient, while other means can fail.

$\Ec$ can be arbitrarily constructed by a \textit{bijective} mapping of the
states; however, for two-band systems, one can always choose Bloch surface,
which is the standard $S_2$ sphere: it is \textbf{universal} in the sense that
it can represent all two-band systems.

In three-band systems, one finds a very short list of surfaces, which can
represent them. In particular, we have proved the existence of a generic surface
$s_6$, which applies for almost all of them and can therefore be considered as
the \textbf{\textit{generalized Bloch surface}} of three-band systems. Its
universality makes it a very powerful device, which can be used to study any
$3\times3$ matrix representation of a hamiltonian.

As we will explain, Bloch surface does not suffice to classify singular
mappings: actually, one must not determinate $\pi_1(\Ec)$ but $\pi_1(\Ec')$
instead, where $\Ec'$ is a specific subspace of $\Ec$, called \textit{effective}
surface, $\Ec' \subset\Ec$. Set $\Ec'$ is associated to the specific symmetries,
i.e.  to the specific band structure of the system.  In other words, the
\textit{universality} of classification surface $\Ec$ does not imply that of
classification groups $\pi_1(\Ec')$.  Conversely, $\pi_1(\Ec')$ is not a
subgroup of $\pi_1(\Ec)$ and must be calculated separately.  Nevertheless, the
universality of $\Ec$ is a very powerful feature since it exactly circumscribes
the possible spaces $\Ec'$, the first homotopy group of which are relevant. This
is true for both two and three-band systems, however, in the three-band case,
the generalized Bloch surface has a very complicated structure with holes, for
which a partial classification of paths can be immediately established
\textit{without} the determination of a specific effective surface
$\mathcal{E}'$, contrary to the two-band case.

In this article, we deal with two and three-band systems. We first detail the
determination of Bloch sphere $S_2$, in a synthetic and pedagogical way. We then
present a complete classification of all three-band cases, revealing essential
differences with that of two-band ones, and give several examples of
application.

\section{Two-band systems}

\subsection{Matrix representation of physical states}

Two-band systems can be represented by $2\times2$ hamiltonian matrices $H$.
Physical states are related to eigenvectors $\vert e,m\rangle$ of $H$ associated
to each energy $e$, where all other degrees of freedom are encoded by symbol
$m$.  In order to get rid of free phase, we will use the representation of
physical states by projectors $\Pi_{e,m}$, which are related to eigenvectors
through $\Pi_{e,m}=\frac{\vert e,m\rangle\langle e,m\vert}{\langle e,m\vert e,m
\rangle}$.  $\Pi_{e,m}$ are $2\times2$ matrices, so one can introduce
$\alpha\in\C$ and $\vec{U}_{e,m}= \Bigg(\!\!\begin{array} {c}x_{e,m}\\[-5pt]
y_{e,m}\\[-5pt]z_{e,m} \end{array}\!\!\Bigg)\in\C^3$ and write
$\Pi_{e,m}={\bm\sigma}.\vec{U}_{m,e}+ \alpha\;I$ where $I$ is identity,
${\bm\sigma}=(\sigma_x,\sigma_y,\sigma_z)$ and $\sigma_i$ are Pauli matrices.

\subsection{Matrix component equations}

In order to get a basis of eigenvectors, which can represent all physical
states, matrices $\Pi_{e,m}$ fulfill two kinds of conditions: inner relations,
that insure each $\Pi_{m,e}$ to be a projector; mutual relations that insure
they represent orthogonal states. Inner relations are written
\begin{equation}
\label{in}
\Pi_{e,m}^\dag=\Pi_{e,m},\quad\Pi_{e,m}\Pi_{e,m}=\Pi_{e,m},\quad
\Tr(\Pi_{e,m})=1,
\end{equation}
while mutual ones
\begin{equation}
\label{out}
\Pi_{e_1,m_1}\Pi_{e_2,m_2}=0\quad \forall e_1\ne e_2.
\end{equation}

(\ref{in}) gives $2\alpha\;{\bm\sigma}.\vec{U}_{e,m}+
(\alpha^2+\Vert\vec{U}_{e,m}\Vert^2)I={\bm\sigma}.\vec{U}_{e,m}+\alpha\;I$ and
$\vec{U}^\ast_{e,m}=\vec{U}_{e,m}$, thus one gets $\vec{U}_{e,m}\in\R$ and
\begin{equation}
\alpha=\frac12\;\&\;\alpha^2+\Vert\vec{U}_{e,m}\Vert^2=\alpha
\iff\alpha=\frac12\;\&\;\Vert\vec{U}_{e,m}\Vert=\frac12.
\label{alph}
\end{equation}
\noindent
Note that $\Tr(\Pi_{e,m})=1$ implies $\alpha=\frac12$ and is thus redundant.

(\ref{out}) gives ${\bm\sigma}.(\frac12(\vec{U}_{e_1,m_1}+\vec{U}_{e_2,m_2})+
\ii\;\vec{U}_{e_1,m_1}\times\vec{U}_{e_2,m_2})+
(\vec{U}_{e_1,m_1}.\vec{U}_{e_2,m_2}+\frac14)I=0$, thus one gets
\begin{equation}
\label{vect}
\vec{U}_{e_1,m_1}.\vec{U}_{e_2,m_2}=-\frac14\;\&\;
\vec{U}_{e_1,m_1}+\vec{U}_{e_2,m_2}=
-2\ii\vec{U}_{e_1,m_1}\times\vec{U}_{e_2,m_2}\,.
\end{equation}

(\ref{alph}) and (\ref{vect}) together give finally
\[
\fbox{$\vec{U}_{e_1,m_1}=-\vec{U}_{e_2,m_2}\,.$}
\]
Therefore, all physical state degrees of freedom are encoded by a single vector
$2\vec{U}_{e,m}$, which belongs to real sphere $S_2$. We have proved that the
universal classification surface for two-band systems is Bloch sphere.

However, as explained before, a specific model can be embedded in a subset of
$S_2$. For instance, for Weyl-Wallace model\cite{Wallace}, which describes
non-magnetic graphene, all physical state degrees of freedom are encoded in
equatorial circle $S_1$ included in $S_2$. $S_1$ is the effective classification
surface of this model.

This example gives $\pi(S_2)=0$, while $\pi(S_1)=\Z$. There is indeed a
topological singularity in Weyl-Wallace model, which lies at each contact point
$P_i$ between energy bands and is characterized by a winding number
$\omega_i\in\Z$.  Indeed, we have written this pedagogical review of two-band
systems in order to emphasise the difference between $\Ec$ and $\Ec'$.
Nevertheless, the existence of a universal classification surface is of major
importance, as we will show now for three-band systems.

\section{Three-band systems}
\label{3bandes}

Three-band systems can be represented by $3\times3$ hamiltonian matrices.  One
needs to find a basis of eigenvectors and we will again represent physical
states by projectors $\Pi_a$ (where $a=(a_i)_{i=1..8}$ is a real vectors in
eight dimensions and plays the same role as $(e,m)$ in the two-band case) which
can be decomposed\cite{Goyal,Abramovici} into eight Gell-Mann matrices
$\lambda_i$ and identity $I$,
\[
\Pi_a=\frac13I+\frac1{\sqrt3}\sum_{i=1}^8 a_i\lambda_i
\]
and still satisfy (\ref{in}) and (\ref{out}). (\ref{in}) now reads
\begin{equation}
\label{ain}
a.a=1\quad\hbox{and}\quad a\star a=a\;,
\end{equation}
where the definition of $\star$ product is recalled in appendix, while
(\ref{out}) becomes
\begin{equation}
\label{aout}
a.b={-}\frac12\quad\hbox{and}\quad a\star b={-}a-b\quad\forall a\ne b\in\R^8\;.
\end{equation}

There is no need to introduce $\Pi_c$, representing a third independent
eigenvector, since one would get $\Pi_a+\Pi_b+\Pi_c=I$.

This system has been solved\cite{Abramovici} in the real case defined by
\renewcommand{\theequation}{$R$} 
\begin{equation}
\label{R}
\fbox{$\alpha_2=\alpha_5=\alpha_7=0$ $\forall\alpha=a,b$.} 
\end{equation}

In the following, we will write $\widehat{\rm eq}$ the nonzero side of any real
algebraic equation (eq), such that it writes $\widehat{\rm eq}=0$, where
$\widehat{\rm eq}$ factorizes in real prime algebraic factors\cite{prime}.  In
addition, writing a variable with index $a$, like $v_a$, means that $v_a$ can be
expressed with $a_1$, ..., $a_8$ components, $v_a=v(a_1,..,a_8)$.  If a variable
must be expressed with both $a$ and $b$ components, we still write $v_a$, thus,
$v_b=v_a\!\rfloor_{\atop\!\!a\leftrightarrow b}$.  Be aware that all equations
in the article are valid when one applies $a\leftrightarrow b$, except the
parametrization ones, so we will omit such exchanged configurations.  For
symmetrical expression, we skip index $a$ which becomes useless since one would
get $v_a=v_b$.  Eventually, the domain of variables $\alpha_i$, with
$\alpha=a,b$, is exactly $-\frac{\sqrt3}2\le\alpha_i\le \frac{\sqrt3}2$
$\;\forall i=1..7$ while $-1\le\alpha_8\le\frac12$.

Here we present the complete general solution, which parts into six different
cases. 

\subsection{First $S_2$ case}

	When $a_8=b_8=\frac12$, one finds $a_i=b_i=0$ $\forall i=4..7$ and $a_i={-}b_i$
$\forall i=1..3$.  Equations (\ref{ain}) and (\ref{aout}) reduce to sphere $s_2$
of equation 
\renewcommand{\theequation}{$S_2$} 
\begin{equation}
\label{S2}
\fbox{${a_1}^2+{a_2}^2+{a_3}^2=\frac34$},
\end{equation}
with 3 degrees of freedom $(a_i)_{i=1..3}$.

\subsection{Second $S_2$ case}

	When $a_8=\frac12$ and $b_8={-}1$, one finds $a_i=b_i=0$ $\forall i=4..7$ and
$b_i=0$ $\forall i=1..3$. Equations (\ref{ain}) and (\ref{aout}) reduce to
sphere $s_2$. 

\subsection{First $S_4$ case}

When $b_8=\frac12$ and $-1{<}a_8{<}\frac12$, one finds $b_i=0$ $\forall
i=4..7$.  (\ref{ain}) and (\ref{aout}) reduce to ellipsoid $s_4$ of equation
\renewcommand{\theequation}{$S_4$} 
\begin{equation}
\label{S4}
\fbox{$\frac43(a_8+\frac14)^2+\sum_{i=4}^7{a_i}^2=\frac34$},
\end{equation}
with 5 degrees of freedom $(a_i)_{i=4..8}$. The parametrization of other
variables becomes $a_i=(-1)^i\frac{\sqtrois\,\ei_{\!a}}{1-2a_8}$ $\forall i=
1,2$, $a_3= \frac{\sqtrois\At_a}{2(1-2a_8)}$ and $b_i=-\frac{3a_i}{2(1+a_8)}$
$\forall i=1..3$, with $\Aa_a={a_4}^2+{a_5}^2$, $\Ab_a={a_6}^2+{a_7}^2$,
$\At_a=\Aa_a-\Ab_a$, $\ea_a=a_4a_6+a_5a_7$ and $\eb_a=a_4a_7-a_5a_6$. 

\subsection{Second $S_4$ case}

When $a_i={-}b_i$ $\forall i=4..7$, one finds $b_i=\frac{1-2a_8}{2(1+a_8)}a_i$
$\forall i=1..3$ and $b_8=-\frac12-a_8$. The parametrization of $a_i$, $i=1..3$,
is unchanged from the previous case contrary to that of $b_i$. Equations
(\ref{ain}) and (\ref{aout}) reduce to ellipsoid $s_4$.

From now on, these four cases will be called \textit{atypical}. Additional
conditions, for instance $a_7=0$ in the first $S_4$ case, give subcases, which
will not be distinguished here, although their equations differ (for instance
$S_3\ne S_4$), and are also \textit{atypical}.

\subsection{General case}

All other solutions can be expressed as the intersection of paraboloid of
equation
\renewcommand{\theequation}{$\gamma$}
\begin{equation}
\label{gamma}
\fbox{$\ga_a=\gb_b$}
\end{equation}
and the 10$^{\rm th}$ degree algebraic curve of equation: 
\renewcommand{\theequation}{\arabic{equation}$_p$} 
\setcounter{equation}{6}
\begin{equation}
(\widetilde{H}-2h)((3+\upsilon)(H+2h)-(A_a-3)\upsilon\, A_b)
=3A_a(H+2h+\upsilon A_b)(A_b+\upsilon)\qquad
\label{eq47}
\end{equation}
where  $A_a=\Aa_a+\Ab_a$, $H=\Aa_a\Aa_b+\Ab_a\Ab_b$, $\widetilde{H}=\Aa_a\Ab_b+
\Aa_b\Ab_a$, $\upsilon=\da+\db$, $\ga_a=a_5b_4-a_4b_5$, $\gb_a=
a_7b_6-a_6b_7$, $\da=a_4b_4+a_5b_5$, $\db=a_6b_6+a_7b_7$ and
$h=\da\db+\ga_a\gb_a$. The explicit expression of (\ref{eq47}) is given in
appendix.

I call (\ref{eq47}) the \textit{parent} equation of forthcoming (\ref{eq7}).
Although (\ref{eq47}) is not symmetrical, we have skipped index $a$ because
(\ref{eq47}) proves to be both the \textit{parent} of (\ref{eq7}) and ($7_b$),
where $(7_b)=(7_a)\!\rfloor_{\atop\!\!a\leftrightarrow b}$;
therefore, one does not need to use another \textit{parent} equation.

The number of parameters in (\ref{eq47}) is 8, but one parameter must be
discarded in order to take (\ref{gamma}) into account. Reducing the number of
parameters of (\ref{eq47}) is done in the following generic case by discarding
$b_7$ thanks to (\ref{gamma}).

\setcounter{equation}{6}
\subsection{Generic case}

In most cases, (\ref{ain}) and (\ref{aout}) reduce to

\renewcommand{\theequation}{\arabic{equation}$_a$} 
\begin{equation}
(t_a^2+3t_aa_6-s_a(A_a-3))(s_aA_a-2a_6t_a\da+u_a{a_6}^2)=3A_a(s_a+a_6t_a)^2\quad
\label{eq7} 
\end{equation}
with $s_a={\ga_a}^2+2\ga_a a_7b_6+\Ab_a{b_6}^2+\Aa_b{a_6}^2$,  $u_a=
2(\da^2+{\ga_a}^2)-\Aa_b\At_a$ and $t_a=a_6\da+a_7\ga_a+b_6\Ab_a$, since
$\frac{a_6\da+a_7\ga_a+b_6\sAb_a}{{a_6}^5}\widehat{7_a}=
\widehat{7_p}\rfloor_{b_7\to\frac{\ga_a+a_7b_6}{a_6}}$. The explicit expression of (\ref{eq7}) is given in appendix.

I call (\ref{eq7}) a \textit{basic} equation; it is universal, meaning that any
non-specific case follows it. Index $a$ in its numbering is similar to that
introduced for variables. 

There are exactly 7 parameters $(a_i)_{i=4..7}$ and $(b_i)_{i=4..6}$ in
(\ref{eq7}). The parametrization of other variables reads $a_1=\frac{\sqtrois\,
\ea_{\!a}q_a}{w_a}$, $a_2={-} \frac{\sqtrois\,\eb_{\!a}q_a}{w_a}$,
$a_3=\frac{\sqtrois\At_aq_a}{2w_a}$, $b_1=\frac{{v_1}_bw_a}{\sqtrois\,t_aq_a}$,
$b_2=\frac{{v_2}_bw_a} {\sqtrois\,t_aq_a}$,
$b_3=\frac{z_bw_a}{2\sqtrois\,a_6t_aq_a}$, $b_7=\frac{\ga_a+a_7b_6}{a_6}$,
$a_8=\frac12-\frac{w_a}{2q_a}$ and $b_8=\frac12+\frac{3t_aq_a}{2a_6w_a}$, where
${v_1}_a=a_5\ga_a-a_6\ca_a$, ${v_2}_a= a_6\cb_a-a_4\ga_a$,
$q_a={\ga_a}^2+{a_6}^2(\Aa_b+\da)+b_6\Ab_a(a_6+b_6)+a_7\ga_a(a_6+2b_6)$, $z_a=
{\ga_a}^2-{b_6}^{\ 2} \Aa_a-2a_6b_7\ga_a+{a_6}^2\Ab_b$, $\ca_a=a_4b_6+a_5b_7$,
$\cb_a=a_4b_7-a_5b_6$ and $w_a={\ga_a}^2(\Aa_a+2{a_6}^2)
-2a_7\ga_a(a_6\da-b_6\Aa_a)+\Ab_{a}({b_6}^{\ 2}\Aa_a+{a_6}^2\Aa_b-2a_6b_6\da)$.

Instead of $b_7$, one could discard any variable $(\alpha_i)$, with $i=4..7$ and
$\alpha=a,b$, still using (\ref{gamma}) in (\ref{eq47}). Altogether, on can get,
through this process, eight different basic equations, which we write
$(e_{i\alpha})$. For example, $(e_{7a})=$ (\ref{eq7}).  For instance, replacing
$b_7$ by $a_7$ gives basic equation $(e_{7b})=(7_b)$ but other changes
are more involved.  

\section{Discussion}

(\ref{eq7}) is a universal equation that describes a universal classification
surface, written $s_6$, spanned in the 7-dimension space of parameters
$\{a_4,a_5,a_6,a_7,b_4,b_5,b_6\}$. Almost any three-band hamiltonian system can
be mapped into $s_6$\cite{hal}. A complete study of its fundamental group would
be an extremely powerful device\cite{F123}; however, for each specific
hamiltonian, it will be much easier to map a path turning around a suspected
singularity into $s_6$ and to reveal indeed a cylindrical hole of the universal
classification surface.  When this occurs, it proves the topological nature of
the singularity and provides the corresponding winding number.

(\ref{eq47}) is the \textit{parent} equation of (\ref{eq7}), it is universal
in the same sense, although it may not be unique. It is also parent of (7$_b$).
The main interest of this parent equation is that it holds in \textbf{all}
cases, but atypical ones, whereas some specific cases are not atypical but do
not follow (\ref{eq7}).

As for the two-band hamiltonian systems, where Bloch sphere is trivial, one must
sometimes investigate the fundamental group $\pi_1(x)$ of a classification
surface $x$, where $x$ is related to the specific basic equation $(x)$ of a
case.  In general, $x\subset s_6$ and $(x)$ is deduced from universal
(\ref{eq7}); in particular cases, $(x)$ is deduced from (\ref{eq47}); in
atypical ones, from (\ref{S2}) or (\ref{S4}) equations.

\section{Applications}

\subsection{Lieb-kagome model}

\begin{figure}[H]
\begin{center}
\includegraphics[width=6cm]{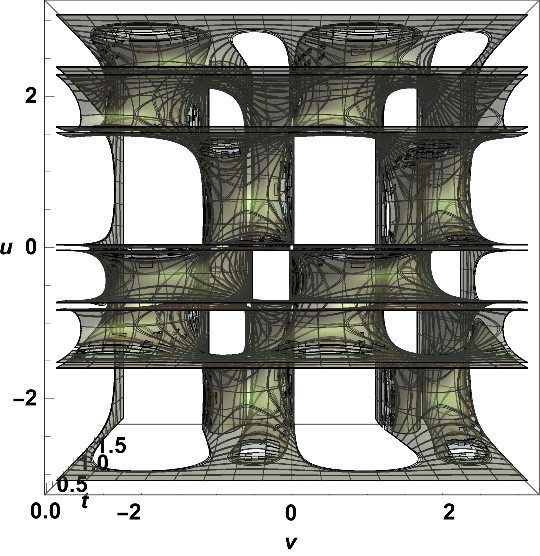}
\end{center}
\vglue-0.5cm
\caption{Representation of $\mathcal{S}$, as explained in
Ref.~\cite{Abramovici}.}
\label{St}
\end{figure}

Lieb-kagome hamiltonian\cite{Parabolic,Lih-King,Abramovici} follows condition
(\ref{R}) and its universal surface is $\mathcal{S}$ (sketched in
Fig.~\ref{St}), the equation of which can be directly deduced from (\ref{eq7}).
Every singularity of this system is mapped into holes of $\mathcal{S}$.
However, some other holes in $\mathcal{S}$ are irrelevant for this model because
the mapping is only injective and not surjective: it does not cover the whole
surface $\mathcal{S}$ but a part of it, which is the \textit{effective}
classification surface.
\vglue-0.5cm

This situation is very general and occurs in many cases. It applies
\textit{mutas mutandis} when the basic equation can only be deduced from
(\ref{eq47}).

\subsection{Real case} 

More generally, $\mathcal{S}$ is the universal classification surface of
\textbf{all} systems, for which (\ref{R}) is fulfilled. However, this does
\textbf{not imply} that, in such systems, the singularities can be directly
classified by $\pi_1(\mathcal{S})$, as it is in the very fortunate case of
Lieb-kagome model. 

\subsection{Generalized Haldane model on Lieb lattice}

We introduce\cite{Doucot} the Bloch Hamiltonian
\[
\Hg=\hbar\pmatrix{
0&\Omega_1\cos\frac{k_x}2&-\ii\Omega_3\cos\frac{k_x-k_y}2\cr
\Omega_1\cos\frac{k_x}2&\du&\Omega_2\cos\frac{k_y}2\cr
\ii\Omega_3\cos\frac{k_x-k_y}2&\Omega_2\cos\frac{k_y}2&\dt\cr}
\]
and will consider the case with $\Omega_1=\Omega_2=\Omega_3=1$. 

\begin{figure}[H]
\begin{center}
\includegraphics[width=7cm]{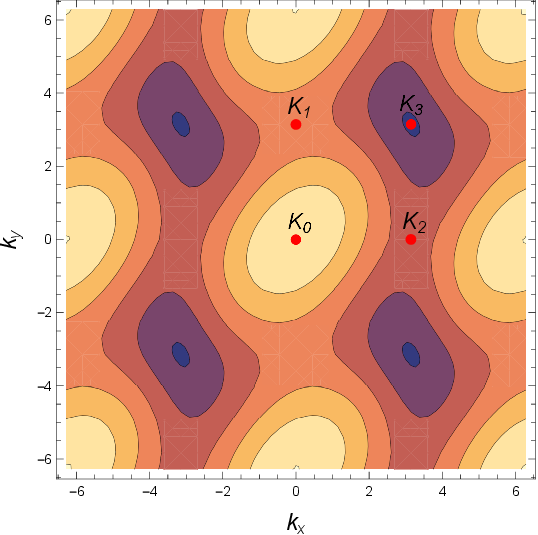}
\end{center}
\caption{Contours of the highest eigenenergy of $\Hg$ with $k_x$ and $k_y$
varying in $[-2\pi,2\pi]$. Discontinuities $K_0$, $K_1$, $K_2$ and $K_3$ are
indicated. Both axis are necessarily $2\pi$-periodic but, according to the
values of $(\du,\dt)$, the diagram can be $\pi$-periodic in $k_x$-direction or
in $k_y$-direction or both or none. Here, $(\du,\dt)=(-\frac12,\frac12)$.}
\label{singH}
\end{figure}

\begin{figure}[H]
\begin{center}
\includegraphics[width=7cm]{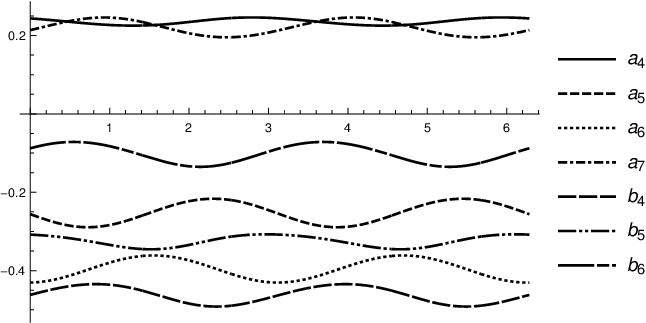}
\end{center}
\caption{Plots of $\alpha_i(\cos t,\sin t)$ for $\alpha=a,b$ and $i=4..7$
(withdrawing $b_7$) for arbitrary values of $(\du,\dt)$, here
$(\du,\dt)=({-}0.5,1.2)$.}
\label{nn00}
\end{figure}

\begin{figure}[H]
\begin{center}
\includegraphics[width=4.5cm]{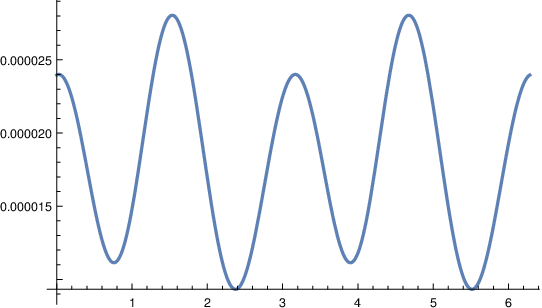}
\end{center}
\caption{Plot of $|\widehat{7_a}(Q_t)|$, the distance from $Q_t$ to $s_6$ for
$t\in[0,2\pi]$ and arbitrary values of $(\du,\dt)$, here
$(\du,\dt)=(0.5,0.3)$.}
\label{dis00}
\end{figure}

The study of eigenvectors of $\Hg$ reveals four singularities in the reciprocal
space, $K_0$ corresponding to $(k_x,k_y)=(0,0)$, $K_1$ to $(0,\pi)$, $K_2$ to
$(\pi,0)$ and $K_3$ to $(\pi,\pi)$, see Fig.~\ref{singH}.  Contrary to
Liev-kagome system, this one is not pathological and energy bands do not
collapse. The four singularities are found from symmetry
considerations\cite{Doucot} and their positions are confirmed independently by
the hereby calculations. Hamiltonian $\Hg$ does not respect symmetry (\ref{R}),
all components of eigenvectors are non-zero. They respect (\ref{eq7}), so one
can analyse its singularities in $s_6$; in this classification space, the system
is faithfully represented by $(a_4,a_5,a_6,a_7,b_4,b_5,b_6)$ and we write
$X(k_x,k_y)= (a_4\skxy,a_5\skxy,a_6\skxy,a_7\skxy,b_4\skxy,b_5\skxy,b_6\skxy)$.
The exploration of singularity $K_0$ proves easy: let $\mathcal{C}_0$ be the
circle described by $(k_x,k_y)=(\cos t,\sin t)$ and turning around $K_0$, one
observes that the trajectory of $P_t=X(\cos t,\sin t)$ in $s_6$ is a loop with
period $\pi$, as shown in Fig.~\ref{nn00}; this means that $\mathcal{C}_0$ maps
into a double loop, so $K_0$ corresponds to winding numbers $\omega_0=\pm2$.
Thus $P_{t+\pi}=P_t$ $\forall t$, which allows us to plot in Fig.~\ref{dis00}
the distance\cite{distance} from $Q_t\equiv(P_t+P_{t+\frac\pi2})/2$ to $s_6$
versus $t$. It is always strictly positive, which proves that $Q_t$, which
describes a continuous loop while $t$ varies from 0 to $\pi$, lies strictly
outside of $s_6$. Since $Q_t$ is the isobarycentre of points $P_t$ and its
half-period translate $P_{t+\frac\pi2}$, $P_t$ turns, while moving with the same
parameter $t$, around $Q_t$, therefore the mapping $P_t$ turns around a hole in
$s_6$. This hole cannot be closed, so this mapping is not
contractible\cite{close}.
The hole defined above allows the determination of $\omega_0$. It is not only
cylindrical: decreasing the radius of $\mathcal{C}_0$ down to zero, one finds
that its shape is a six-dimensional multiple cone, which is pinched into a point
$D_0$ at its centre, $D_0$ is the image of $K_0$. More generally, $\forall
i=0,..,3$, one finds that all singularities $K_i$ can be classified by a
multiple cone in $s_6$, pinched into a point $D_i$, which is the image of $K_i$.

\begin{figure}[H]
\begin{center}
\includegraphics[width=7.5cm]{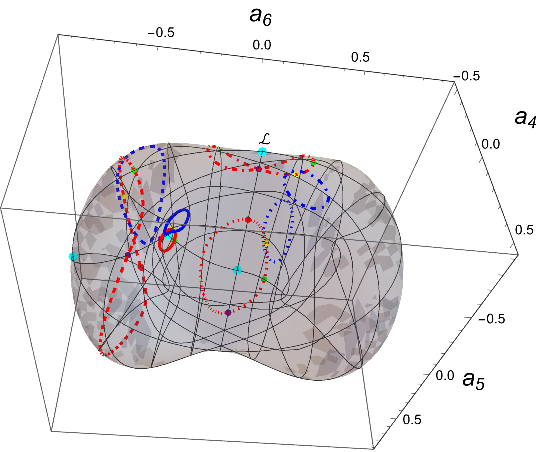}
\end{center}
\caption{Representation of the mapping by $\tau_1$ of several circles of
$(k_x,k_y)$-plane. $\tau_1(s_6)$ is pinched along a line $\mathcal{L}$, which is
indicated in the middle of the figure. Large cyan points represent $\tau_1(D_i)$
for $i=1..4$, mappings are in red when the circle is around a singularity, in
blue when it avoids it. Red, yellow, green and purple points follow this order
in non-trivial paths but may overlap. Plain lines correspond to singularity
$K_0$ ($\tau_1(D_0)$ is a singularity of the surface, as shown in
Fig.~\ref{det00}).  Dotted lines correspond to singularity $K_1$; $\tau_1(D_1)$
is at the middle of $\mathcal{L}$ and the non-trivial path crosses $\mathcal{L}$
twice.  Dashed lines correspond to singularity $K_2$ (the 8-shape is
artificially created by the two-dimensional projection of the drawing).
Dot-dashed lines correspond to singularity $K_3$, one observes that the
non-trivial one makes a loop with 8-shape, which collapses at one extremity of
$\mathcal{L}$.}
\label{pr456x}
\end{figure}

\begin{figure}[H]
\begin{center}
\includegraphics[width=5cm]{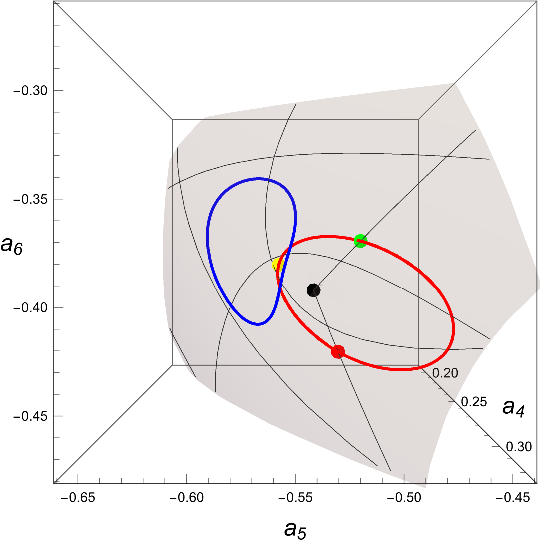}
\end{center}
\caption{Mapping of two circles in $(k_x,k_y)$-plane, the first one turns around
$K_0$ and maps into a loop turning twice around the singularity figured by a
black point. Colored points follow the same order as in the previous figure. The
second circle avoids $K_0$ and maps into a single loop, turning once and
avoiding the black point, it is therefore trivial.}
\label{det00}
\end{figure}

For $(\du,\dt)= (\frac12,\frac12)$, $P_t$ turns around
$(0.292,-0.52,-0.4,0.27,-0.58,0.1,-0.2)$.  This determination depends on
$(\du,\dt)$ and on the singularity $K_i$, with $i=0..3$. The situation is very
intricate for cases $i\ne0$, so we will only study the hole corresponding to
singularity $K_0$.  Also, the directions of the hole should be studied
separately.  One can map $P_t$ in $\tau_1(s_6)$, the projection of $s_6$ in
$(a_4,a_5,a_6)$ coordinates: $\tau_1(P_t)$ makes a double loop around a
singularity of $\tau_1(s_6)$, as shown in Fig.~\ref{pr456x}.  In order to see
the mapping of $\mathcal{C}_0$ more clearly, we have made a zoom in
Fig~\ref{det00}.

\begin{figure}[H]
\begin{center}
\includegraphics[width=7cm]{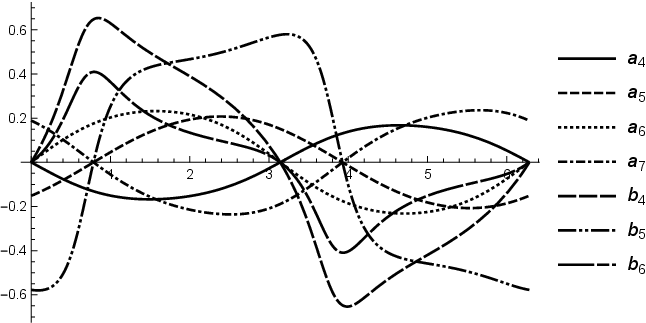}
\end{center}
\caption{Plots of $\alpha_i(\cos t,\pi+\sin t)$ for $\alpha=a,b$ and $i=4..7$
(withdrawing $b_7$) for arbitrary values of $(\du,\dt)$, here
$(\du,\dt)=({-}0.5,0.9)$.}
\label{nn0pi}
\end{figure}

\begin{figure}[H]
\begin{center}
\includegraphics[width=5cm]{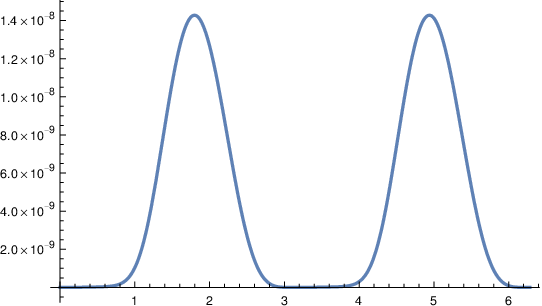}
\end{center}
\caption{Plot of $|\widehat{7_a}(S^1_t)|$, the distance from $S^1_t$ to $s_6$
for $t\in[0,2\pi]$, with $\theta=1$ and arbitrary values of $(\du,\dt)$,
here $(\du,\dt)=(0.5,0.3)$.}
\label{dis0pi}
\end{figure}

\begin{figure}[H]
\begin{center}
\includegraphics[width=4cm]{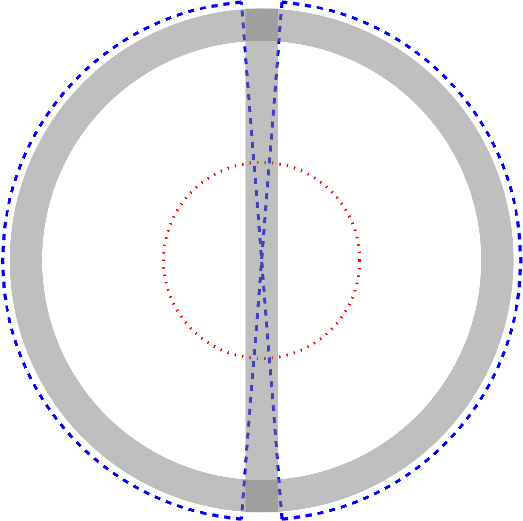}
\end{center}
\caption{Sketch of the orthogonal section of the hole structure, corresponding
to singularity $K_2$. The dashed line stands for $R_t$, one observes that the
hole divides into two separate holes, which frontier is flat (at least in some
dimensions). The dotted line stands for $S^1_t$.}
\label{schema}
\end{figure}

Let us move to the analysis of singularity $K_1$, which analysis  proves much
more involved than that of $K_0$. Let $\mathcal{C}_1$ be the circle described
by $(k_x,k_y)=(\cos t,\pi+\sin t)$ and turning around $K_1$, one observes in
Fig.~\ref{pr456x} the projection of $R_t=X(\cos t,\pi+\sin t)$ in $\tau_1(s_6)$.
Loop $R_t$ is simple, this is confirmed in Fig.~\ref{nn0pi}.
Taking advantage of this, we plot the distance from $S^\theta_t\equiv
\cos(\theta)^2R_t+\sin(\theta)^2R_{t+\pi}$ to $s_6$ versus $t$ in
Fig.~\ref{dis0pi}. Almost all values of $\theta$ can be chosen, giving a
non-zero distance.
However, this distance is zero for $t=0$ and $t=\pi$. Moreover, this plot gives
a constant zero distance when $\theta=\frac\pi4$. This can be correlated to the
crossing of line $\mathcal{L}$ observed in Fig.~\ref{pr456x} and interpreted
as a more complicated hole structure in $s_6$, which can be schematized by its
orthogonal section in Fig.~\ref{schema}. This sketch provides indeed the
configuration, for which the barycentre $S^\theta_t$ (constructed the same way
than $Q_t$, taking into account the doubling of period and choosing weights
$\cos(\theta)^2$ and $\sin(\theta)^2$ instead of uniform weights $\frac12$)
joins surface $s_6$ twice, while $S^{\pi/4}_t$ always lies in $s_6$.
The non triviality of $R_t$ is established by the two holes, with the same
confidence than that of $P_t$. Altogether, we have established that $K_1$
corresponds to winding numbers $\omega_1=\pm1$.

\begin{figure}[H]
\begin{center}
\includegraphics[width=7cm]{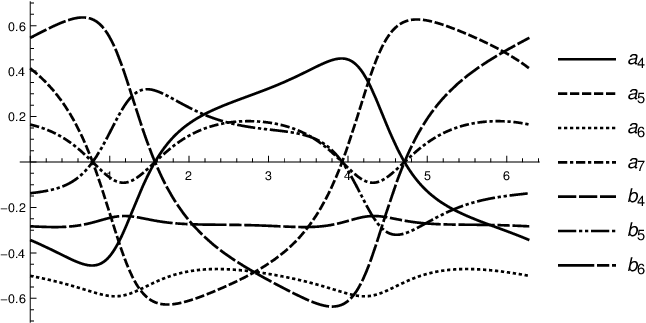}
\end{center}
\caption{Plots of $\alpha_i(\pi+\cos t,\sin t)$ for $\alpha=a,b$ and $i=4..7$
(withdrawing $b_7$) for arbitrary values of $(\du,\dt)$, here
$(\du,\dt)=(0.5,0.9)$.}
\label{nnpi0}
\end{figure}

\begin{figure}[H]
\begin{center}
\includegraphics[width=5cm]{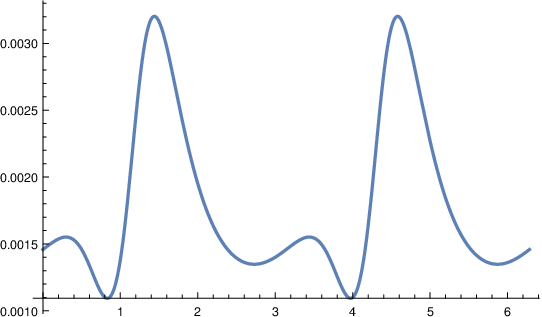}
\end{center}
\caption{Plot of $|\widehat{7_a}(U_t)|$, the distance from $U_t$ to $s_6$
for $t\in[0,2\pi]$ and arbitrary values of $(\du,\dt)$, here
$(\du,\dt)=(0.5,0.3)$.}
\label{dispi0}
\end{figure}

Let us move to the analysis of singularity $K_2$. Let $\mathcal{C}_2$ be the
circle described by $(k_x,k_y)=(\pi+\cos t,\sin t)$ and turning around $K_2$,
one observes in Fig.~\ref{pr456x} the projection of $T_t=X(\pi+\cos t,\sin t)$
in $\tau_1(s_6)$. Loop $T_t$ is simple, this is confirmed in 
Fig.~\ref{nnpi0}.
Taking advantage of this, we plot the distance from $U_t\equiv(T_t+T_{t+\pi})/2$
to $s_6$ versus $t$ in Fig.~\ref{dispi0}. It is always strictly positive and
indicates that the mapping $T_t$ turns around a hole in $s_6$ (using the same
argument used for $P_t$, while taking into account the doubling of period).  The
non triviality of $T_t$ is established with the same confidence than that of
$P_t$.  Altogether, we have established that $K_2$ corresponds to winding
numbers $\omega_2=\pm1$.

\begin{figure}[H]
\begin{center}
\includegraphics[width=7cm]{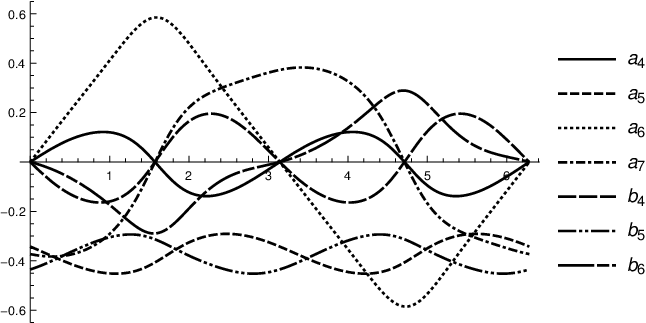}
\end{center}
\caption{Plots of $\alpha_i(\pi+\cos t,\pi+\sin t)$ for $\alpha=a,b$ and
$i=4..7$ (withdrawing $b_7$) for arbitrary values of $(\du,\dt)$, here
$(\du,\dt)=({-}0.5,0.9)$.}
\label{nnpipi}
\end{figure}

\begin{figure}[H]
\begin{center}
\includegraphics[width=5cm]{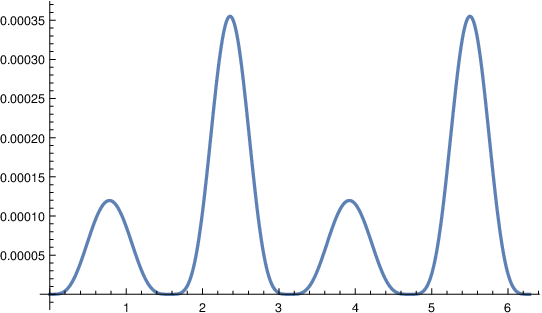}
\end{center}
\caption{Plot of $|\widehat{7_a}(W_t)|$, the distance from $W_t$ to $s_6$
for $t\in[0,2\pi]$, and arbitrary values of $(\du,\dt)$,
here $(\du,\dt)=(0.5,0.3)$.}
\label{dispipi}
\end{figure}

\begin{figure}[H]
\begin{center}
\includegraphics[width=4cm]{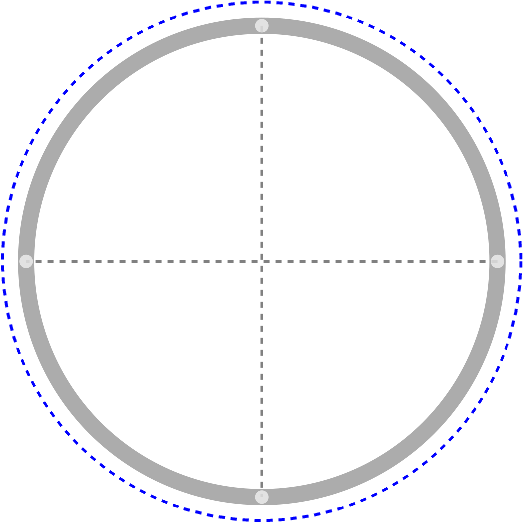}
\end{center}
\caption{Sketch of the orthogonal section of the hole structure, represented by
a circle with radius $r$, corresponding to singularity $K_3$. The blue dashed
line (outside of the plain circle) stands for $V_t$, the dashed line between
points $(0,r)$ and $(0,-r)$ represents an attachment, these points are merged
into a unique one, and so are points $(r,0)$ and $(-r,0)$.}
\label{schena}
\end{figure}

Let us move to the analysis of singularity $K_3$. Let $\mathcal{C}_3$ be the
circle described by $(k_x,k_y)=(\pi+\cos t,\pi+\sin t)$ and turning around
$K_2$, one observes in Fig.~\ref{pr456x} the projection of $V_t=X(\pi+\cos t,
\pi+\sin t)$ in $\tau_1(s_6)$. Loop $V_t$ is simple, this is confirmed
in Fig.~\ref{nnpipi}.
Taking advantage of this, we plot the distance from $W_t\equiv
(V_t+V_{t+\pi})/2$ to $s_6$ versus $t$ in Fig.~\ref{dispipi}.
This distance is always strictly positive, except for $t=n\frac\pi2$ $\forall
n\in\Z$ at which points it is zero. In order to interpret the structure, we
have also checked that $W_t^\theta$ is distant of $s_6$ $\forall
\theta\in[0,2\pi[$ and $\forall t\ne n\frac\pi2$ with $n\in\Z$ ($W_t^\theta$ is
defined exactly as $S_t^\theta$). This can be correlated to the 8-shaped
observed in Fig.~\ref{pr456x} and interpreted as a more complicated hole
structure in $s_6$, which can be schematized by its orthogonal section in
Fig.~\ref{schena}. This sketch provides indeed the configuration, for which the
isobarycentre $W_t$ (constructed the same way than $Q_t$, taking into account
the doubling of period) joins surface $s_6$ four times.
The non triviality of $V_t$ is established by the hole sketched in
Fig.~\ref{schena}, with the same confidence than that of $P_t$. Altogether, we
have established that $K_3$ corresponds to winding numbers $\omega_3=\pm1$.

We would like to address now the question of the sign $s^i(\du,\dt)$ of
$\omega_i$. Its determination $\sr^i(\du,\dt)$ through a
classification surface like $s_6$ is only relative and a global sign
$\varepsilon$ must be defined \textit{elsewhere}, such that
$s^i(\du,\dt)=\varepsilon \sr^i(\du,\dt)$. For singularity
$K_0$, $\sr^0(\du,\dt)=1$ is trivial, as seen by comparing the
turning directions of all path $\tau_1(P_t)$ when $(\du,\dt)$ varies.
This is confirmed by analysing in detail all copies of Fig.~\ref{nn00} for
various values of $(\du,\dt)$.

\begin{figure}[H]
\begin{center}
\includegraphics[width=8cm]{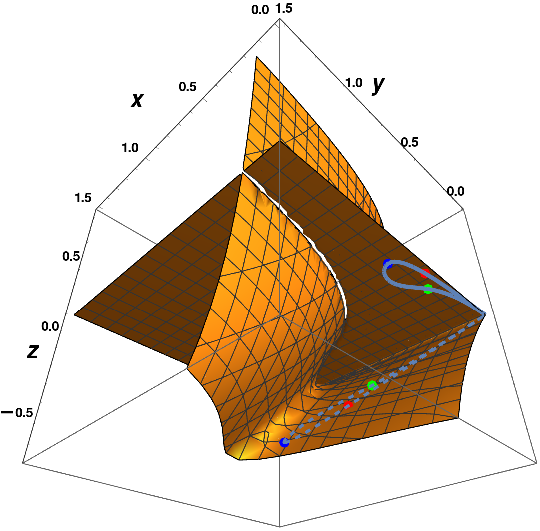}
\end{center}
\caption{Representation of $s_3$, putting $w=\frac9{16}$, its maximal value.
From the definitions of $\alpha_i$ components, with $\alpha=a,b$ and $i=4..7$,
$x$ and $y$ range in $[0,\frac32]$ while $z$ ranges in $[-\frac34,\frac34]$. We
represent two paths obtained by the $\Pim$ projection, the dashed line with
$(\du,\dt)=(\frac12,-\frac12)$, the plain one with $(\du,\dt)=
(-\frac12,\frac12)$, the mappings follow the order $\color{blue}
\bullet\color{red}\bullet\color{green}\bullet$. They are close to different
folds, with the same apparent orientation, although, moving one path
continuously to the orther would give the opposite orientation.}
\label{funi}
\end{figure}

\begin{figure}[H]
\begin{center}
\includegraphics[width=7cm]{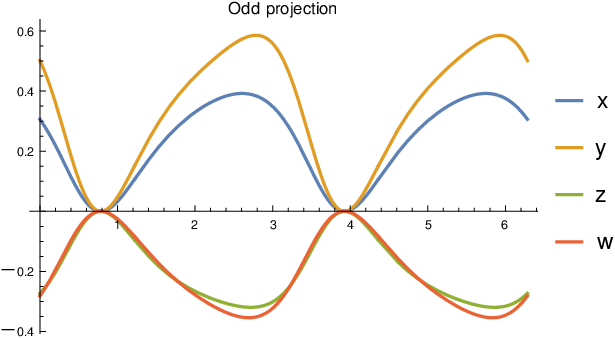}
\includegraphics[width=7cm]{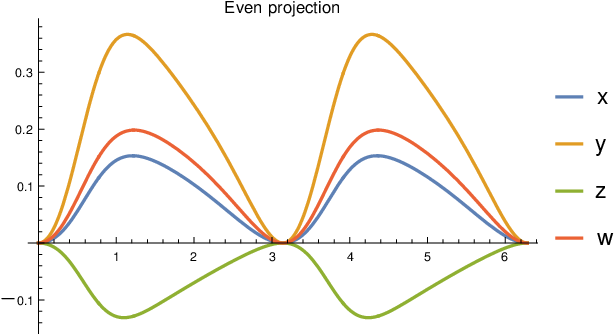}
\end{center}
\caption{Plot of $\Pim(R_t)$ versus $t$ (left) and $\Pip(R_t)$ versus $t$
(right) for $t\in[0,2\pi]$ and arbitrary values of $(\du,\dt)$, here
$(\du,\dt)=(0.5,0.3)$. Be careful that definitions of $(x,y,z,w)$ are different
for $\Pim$ and $\Pip$.}
\label{piper}
\end{figure}

\begin{figure}[H]
\begin{center}
\includegraphics[width=5cm]{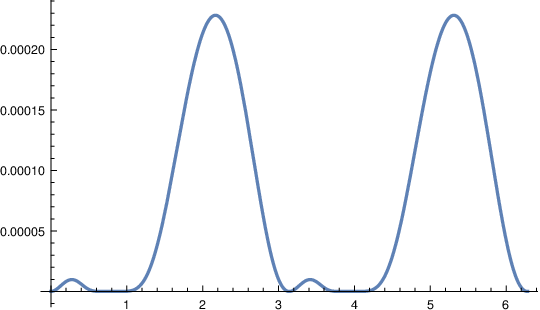}
\includegraphics[width=5cm]{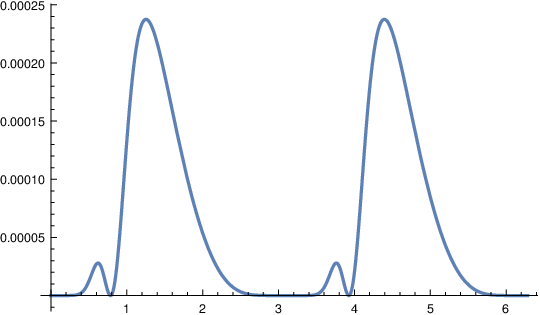}
\end{center}
\caption{Plot of $\Pim(\widehat{\hbox{\ref{univ}}}(R_t))$ versus $t$ (left) and
$\Pip(\widehat{\hbox{\ref{univ}}}(R_t))$ versus $t$ (right) for $t\in[0,2\pi]$
and arbitrary values of $(\du,\dt)$, here $(\du,\dt)=(0.5,-0.9)$. Be careful
that definitions of $(x,y,z,w)$ are different for $\Pim$ and $\Pip$.}
\label{s3pip}
\end{figure}

\begin{figure}[H]
\begin{center}
\includegraphics[width=7cm]{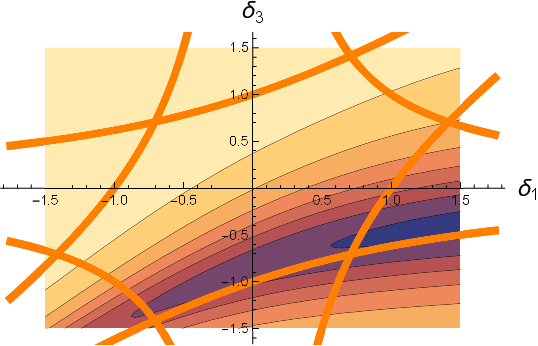}
\end{center}
\caption{Contours of $a_5\skxy b_5\skxy+a_7\skxy b_7\skxy$, with $t=0.2$ and
$(k_x,k_y)=(\cos t,\pi+\sin t)$. $\du$ and $\dt$ vary in $[-\frac32,\frac32]$
and the frontiers of Fig.~\ref{Hg} are indicated.}
\label{chS}
\end{figure}

\begin{figure}[H]
\begin{center}
\includegraphics[width=7cm]{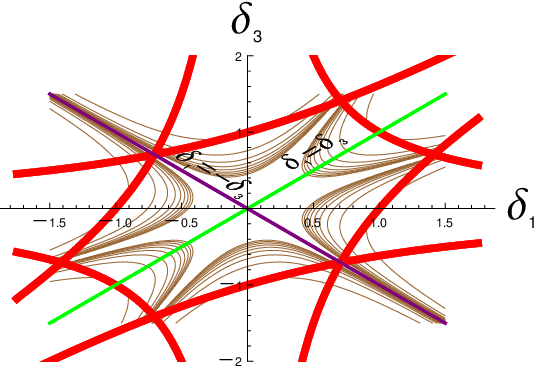}
\end{center}
\caption{Frontiers in the map of winding parameters versus $(\du,\dt)$. The
first and second bisectors are also indicated.}
\label{Hg}
\end{figure}

For singularity $K_1$, a change of sign is manifest at the bottom frontier drawn
in Fig.\ref{Hg} but we could not get to a definitive proof. A huge
inconvenient of the representation chosen in Fig.~\ref{pr456x} is that the
surface on which all circles are mapped is $(\du,\dt)$ dependant. There  exist
universal representations onto which one could map all projections but we could
not achieve their determination yet\cite{deter}.  Instead, we have found that
both projections $\Pim$ and $\Pip$, defined by putting, respectively,
$(a_4,a_6,b_4,b_6)\to(0,0,0,0)$ and $(a_5,a_7,b_5,b_7)\to(0,0,0,0)$ in
(\ref{eq47}), gives the \textbf{same} universal equation
\renewcommand{\theequation}{\arabic{equation}} 
\begin{equation}
\label{univ}
3x\,z(y+z)^2=z\,w(y(3-x)+z(3+z))\,.
\end{equation}
More precisely, $\Pim(\widehat{\hbox{\ref{eq47}}})=0$ gives (\ref{univ}) with
$x={a_5}^2+{a_7}^2$, $y={b_5}^2+{b_7}^2$, $z=a_5b_5+a_7b_7$ and $w=
(a_7b_5-a_5b_7)^2$, while $\Pip(\widehat{\hbox{\ref{eq47}}})=0$ gives
(\ref{univ}) with $x={a_4}^2+{a_6}^2$, $y={b_4}^2+{b_6}^2$, $z=a_4b_4+a_6b_6$
and $w=(a_6b_4-a_4b_6)^2$. We write $s_3$ the surface corresponding to
(\ref{univ}) and show a 3-dimensional representation of $s_3$ in
Fig.~\ref{funi}.
$\Pim(R_t)$ and $\Pip(R_t)$ are found $\pi$-periodic, as shown in
Fig.~\ref{piper}.
This projections are not canonical, indeed neither $\Pim(R_t)$ nor $\Pip(R_t)$
do map exactly on (\ref{univ}). However, by chance, they are close to it.
Actually $\{\Pim(R_0),\Pip(R_0),\Pim(R_{\frac\pi2}),\Pip(R_{\frac\pi2})\}\subset
s_3$, which one shows by plotting $\Pim(\widehat{\hbox{\ref{univ}}}(R_t))$ and
$\Pip(\widehat{\hbox{\ref{univ}}} (R_t))$, in Fig.~\ref{s3pip}. One finds two
close leaves in Fig.~\ref{funi}, which are inversely orientated, from a
topological point of view. Each path moves from one leaf to the other, keeping
the same apparent orientation, when $(\du,\dt)$ approaches the bottom frontier
line of Fig.~\ref{Hg}. So, their effective orientation must be reversed when
crossing this line.
It seems that $z$ drives the change of sign of the orientation. This is
confirmed by the contours of $z=a_5b_5+a_7b_7$ which fits one
of the frontiers shown in Fig.~\ref{chS}.

We must now compare our results to those in \cite{Doucot}. In Fig.~\ref{Hg}, we
show the frontiers that have been found. We added the contours of
$\cot(\theta)=\sqrt{4p(\du,\dt)^3/q(\du,\dt)^2-1}$, with $p(x,y)=
9+x^2-xy+y^2$ and $q(x,y)=2x^3-3x^2y-3x\,y^2+2y^3$, which is related to an angle
$\theta$, the cotangent of which fits exactly these frontiers. $\theta$ is
directly related to the eigenenergies of this model, which are
$\frac{\du+\dt}2+\frac23\sqrt{p(\du,\dt)}\cos(\frac{\theta+\varepsilon\pi}3)$
with $\varepsilon=-1,0,1$.

In \cite{Doucot}, the authors find a winding number $\omega=\pm4$, which sign
changes in the different areas designed by the frontiers of Fig.~.\ref{Hg}. The
combination of two windings $\pm1$ can give an effective winding of $\pm2$ and,
similarly, the combination of two windings $\pm2$ can give an effective winding
of $\pm4$; therefore, some combinations of $\omega_0$, $\omega_1$, $\omega_2$ or
$\omega_3$ give indeed an effective winding number of $\pm4$. On the other hand,
classifying singularities through surface $s_6$ gives exhaustively all primary
winding numbers, so $\omega$ \textbf{must} be some of these combinations.
Moreover, since $\omega$ has different signs in the different areas, described
in Fig.~\ref{Hg}, the related changes of sign must inherit from that of some
$\omega_i$, $i=0..3$; in particular, this reinforces our interpretation of a
sign change for $\omega_1$. But this determination remains currently
questionable and we have renounced to study the signs of $\omega_2$ and
$\omega_3$. Eventually, one must find a more robust method to solve this very
interesting question, from which it will be possible to deduce the relations
between $\omega$ and the other $\omega_i$, $i=0..3$.

Eventually, we deal with an atypical case in the following.

\subsection{Lieb model}

	This model\cite{Lieb,Goldman,Tsai} is extraordinary for several reasons. Its
classification surface is circle $S_1$ and the corresponding basic equation is
thus atypical. It corresponds to Lieb-kagome parameter $t'=0$ and separates from
the general Lieb-kagome surface $\mathcal{S}$, which holds when $0<t'\le1$;
fortunately, we could find surface $\widetilde{\mathcal{S}_1}$, which is valid
both in Lieb and Lieb-kagome cases\cite{Abramovici,validite}, i.e. $\forall
t'\in[0,1]$. Investigating loops while varying continuously $t'$ allows one to
understand why winding numbers tend to $\pm2$ when $t'\to0$, though the exact
$t'=0$ limit, calculated in $S_1$, is $\pm1$: $\widetilde{\mathcal{S}_1}$
consists essentially in two planes, into which each path around a singularity
makes a single loop (which by definition corresponds to winding number 1).  When
the limit $t'=0$ is reached, these planes merge, so that winding numbers fuse
and do not add; thus, this apparent anomaly is explained.

This example seems to indicate that atypical cases arise when the system follows
additional symmetries. In the Lieb case, there is indeed a three-fold degeneracy
of eigenvalues, which is very exceptional.

\section{Conclusion}

It is wonderful that the topological singularities of almost any three-band
hamiltonian can be mapped onto the same universal surface $s_6$. Although one is
not assured to characterize winding numbers in this surface, it contains all
subsurfaces in which they can be defined. Some cases, not following (\ref{eq7}),
do follow another $(e_{i\alpha})$ but, as we believe, it is more efficient to
establish a unique couple of (basic,parent) equations: eventually, one gets only
\textbf{four} different equations (\ref{eq7}), (\ref{eq47}), (\ref{S2}) and
(\ref{S4}) which cover all cases\cite{detail}.

This extends a similar result in real case: $\mathcal{S}$ is universal for
almost all three-band hamiltonian respecting (\ref{R}), but its structure is
much easier to investigate from Fig.~\ref{St}. 

It is not known yet, whether atypical cases have physical applications, but one
can observe that the fourth one (for which general parametrization rules
\ref{RR}, defined in appendix, hold) exactly generalizes the two-band
unique solution.

The way universal classification surfaces are constructed seems to exclude the
influence of each hamiltonian properties in the determination of singularities
but this is a wrong interpretation. Hamiltonians directly govern the way paths
are constructed in $s_6$. Also, their symmetries are responsible for the
reduction from $s_6$ to their effective classification surface. However, in
the Lieb-kagome example, it is true that winding numbers can be immediately
determined in $\mathcal{S}$ and probably in $s_6$ too.

A further simple investigation shall be to study the mapping of paths, defined
in reciprocal space for Lieb-kagome model, into $s_6$, with the hope to
determine a part of its fundamental group.

Eventually, one observes that (\ref{gamma}) is true in any case. For systems
following (\ref{R}) condition, (\ref{gamma}) becomes trivial. We have no
physical interpretation of this condition yet and it will be very interesting to
study it for specific hamiltonians not following (\ref{R}) as $\Hg$.

%\appendix
\section*{Appendix}

\subsection{Definition of the $\star$ product}

Let $a=(a_i)_{i=1..8}$ and $b=(b_i)_{i=1..8}$, two vectors of the $\R^8$ space,
the star product $a\star b$ is defined\cite{Goyal,Abramovici} by $a\star b=
\sum_{i=1}^8 a^j b^k d_{ijk}$ with
\begin{eqnarray*}
&&d_{ii8}=d_{i8i}=d_{8ii}=\frac1{\sqrt3}\qquad \forall i=1,2,3\;;\\ 
&&d_{ii8}=d_{i8i}=d_{8ii}={-}\frac1{2\sqrt3}\quad \forall i=4,..,7\;;\\ 
&&d_{888}={-}\frac1{\sqrt3}\;;\\
&&d_{146}=d_{461}=d_{614}=d_{164}=d_{641}=d_{416}=\frac12\;;\\
&&d_{157}=d_{571}=d_{715}=d_{175}=d_{751}=d_{517}=\frac12\;;\\
&&d_{256}=d_{562}=d_{625}=d_{265}=d_{652}=d_{526}=\frac12\;;\\
&&d_{344}=d_{434}=d_{443}=d_{355}=d_{535}=d_{553}=\frac12\;;\\
&&d_{247}=d_{472}=d_{724}=d_{274}=d_{742}=d_{427}={-}\frac12\;;\\
&&d_{366}=d_{636}=d_{663}=d_{377}=d_{737}=d_{773}={-}\frac12\;.
\end{eqnarray*}

\subsection{Explicit expressions of universal surface $s_6$}
 
The explicit expression of (\ref{eq47}) is
\begin{eqnarray*}
({a_4}^2+{a_5}^2+{a_6}^2+{a_7}^2)(a_4b_4+{b_4}^2+a_5b_5+{b_5}^2+a_6b_6+
{b_6}^2+a_7b_7+{b_7}^2)\times\\
\!\!\!\!\!\!\!\!\!\!\!\!\!\!\!\!
\Bigg(({a_4}^2+{a_5}^2)({b_4}^2+{b_5}^2)+({a_6}^2+{a_7}^2)({b_6}^2+{b_7}^2)
+(a_4b_4+a_5b_5+a_6b_6+a_7b_7)\times\\
\!\!\!\!\!\!\!\!\!\!\!\!\!\!\!\!
\!\!\!\!\!\!\!\!
({b_4}^2+{b_5}^2+{b_6}^2+{b_7}^2)+2\bigg((a_5b_4-a_4b_5)
(a_7b_6-a_6b_7)+(a_4b_4+a_5b_5)(a_6b_6+a_7b_7)\bigg)\Bigg)\\
=\Bigg(({a_6}^2+{a_7}^2)({b_4}^2+{b_5}^2)+({a_4}^2+{a_5}^2)({b_6}^2+{b_7}^2)\\
-2\bigg((a_5b_4-a_4b_5)(a_7b_6-a_6b_7)
+(a_4b_4+a_5b_5)(a_6b_6+a_7b_7)\bigg)\Bigg)\times\\
\!\!\!\!\!\!\!\!\!\!\!\!\!\!\!\!
\Bigg[(3+a_4b_4+a_5b_5+a_6b_6+a_7b_7)\times
\Bigg(({a_4}^2+{a_5}^2)({b_4}^2+{b_5}^2)+({a_6}^2+{a_7}^2)({b_6}^2+{b_7}^2)\\
+2\bigg((a_5b_4-a_4b_5)(a_7b_6-a_6b_7)+(a_4b_4+a_5b_5)(a_6b_6+a_7b_7)\bigg)
\Bigg)\\
\!\!\!\!\!\!\!\!\!\!\!\!\!\!\!\!
-\bigg((-3+{a_4}^2+{a_5}^2+{a_6}^2+{a_7}^2)(a_4b_4+a_5b_5+a_6b_6+a_7b_7)
({b_4}^2+{b_5}^2+{b_6}^2+{b_7}^2)\bigg)\Bigg].
\end{eqnarray*}
The explicit expression of (\ref{eq7}) and thus the equation of $s_6$ is
\begin{eqnarray*}
\!\!\!\!\!\!\!\!\!\!\!\!\!\!\!\!
3({a_4}^2+{a_5}^2+{a_6}^2+{a_7}^2)\Bigg({a_6}^2\bigg(b_4(a_4+b_4)+b_5(a_5+b_5)
\bigg)+({a_6}^2+{a_7}^2)b_6(a_6+b_6)\\
+(a_5b_4-a_4b_5)(a_6a_7+a_5b_4-a_4b_5+2a_7b_6)\Bigg)^2\\
\!\!\!\!\!\!\!\!\!\!\!\!\!\!\!\!
\!\!\!\!\!\!\!\!\!\!\!\!\!\!\!\!
=\Bigg(({a_4}^2+{a_5}^2+2{a_6}^2)(a_5b_4-a_4b_5)^2
-2a_7(-a_5b_4+a_4b_5)\bigg(({a_4}^2+{a_5}^2)b_6-a_6(a_4b_4+a_5b_5)\bigg)\\
+({a_6}^2+{a_7}^2)\bigg({a_6}^2({b_4}^2+{b_5}^2)
-2a_6(a_4b_4+a_5b_5)b_6+({a_4}^2+{a_5}^2) {b_6}^2\bigg)\Bigg)\times\\
\Bigg(\bigg(a_4a_6b_4+a_5a_7b_4+a_5a_6b_5-a_4a_7b_5+({a_6}^2+{a_7}^2)b_6\bigg)
\times\\
\bigg(a_6(3+a_4b_4+a_5b_5)
+{a_6}^2b_6+a_7(a_5b_4-a_4b_5+a_7b_6)\bigg)\\
\!\!\!\!\!\!\!\!\!\!\!\!\!\!\!\!
-(-3+{a_4}^2+{a_5}^2+{a_6}^2+{a_7}^2)
\bigg((a_5b_4-a_4b_5+a_7b_6)^2+{a_6}^2({b_4}^2+{b_5}^2+{b_6}^2)\bigg)\Bigg).
\end{eqnarray*}

\subsection{Subcases} 

We study what happens when $a_i$ follow
additional conditions in more details, through some examples, but we exclude
atypical cases.

Let's consider a system obeying additional conditions $\ca_a=0=
a_4b_i+a_ib_4$, $i=6,7$. Its basic equation becomes $(x_1)$
$3\widetilde{A}_a({\Ab_a}^2
(a_4-b_4)^2-{\Aa_a}^2(a_4+b_4)^2)=16{b_4}^2{\Aa_a}^2{\Ab_a}^2$ and follows basic
equation (\ref{eq7}) since $\widehat{x_1}=\frac{{a_4}^4}{{a_6}^4{b_4}^2}
\widehat{7_a}\rfloor_{b_5\to\frac{b_4a_5}{a_4},b_6\to-\frac{b_4a_6}
{a_4},b_7\to-\frac{b_4a_7}{a_4}}$. There are 5 degrees of freedom
$(a_i)_{i=4..7}$ and $b_4$ and the parametrization of other ones reads
$a_1=-\frac{4b_4\sAa_a\sAb_a\ea_{\!a}}{\sqtrois\At_a x_a}$,
$a_2=-\frac{\eb_{\!a}}{\ea_{\!a}}a_1$, $a_3=-\frac{2b_4\sAa_a\sAb_a}{\sqtrois\,
x_a}$, $b_i=-\frac{{b_4}^2}{{a_4}^2}a_i$ $\forall i=1..3$,
$b_i=-\frac{b_4}{a_4}a_i$ $\forall i=6,7$, $b_5=\frac{b_4a_5} {a_4}$,
$a_8=-\frac12+\frac{3((a_4+b_4){\sAa_a}^2-(a_4-b_4){\sAb_a}^2)}
{8b_4\sAa_a\sAb_a}$ and $b_8=\frac12-\frac{2{b_4}^2\sAa_a\sAb_a}{a_4x_a}$.

If condition $a_6=b_6=0$ is added to the previous condition, the basic equation
of the system becomes $(x_2)$
$3\As_a({a_7}^2(a_4-b_4)^2-{\Aa_a}^2(a_4+b_4)^2)=16{a_7}^4{b_4}^2{\Aa_a}^2$ with
$\widehat{x_2}=\frac{{a_4}^5}{{b_4}^3\sAs_a}\widehat{7_p}
\rfloor_{b_5\to\frac{b_4a_5}{a_4},a_6\to0,b_6\to0,b_7\to-\frac{b_4a_7}{a_4}}$,
where $\As_a=\Aa_a-{a_7}^2$, and does not follow basic equation (\ref{eq7}) but
its parent equation (\ref{eq47}). There are 4 degrees of freedom
$(a_4,a_5,a_7,b_4)$ and the parametrization of other ones reads $a_1=
\frac{\sqtrois a_5y_a} {4a_7b_4\sAa_a}$, $a_2=-\frac{\sqtrois a_4y_a}
{4a_7b_4\sAa_a}$, $a_3=\frac{\sqtrois\sAs_ay_a} {8{a_7}^2b_4\sAa_a}$,
$b_1=\frac{4a_5{a_7}^3{b_4}^2\sAa_a}{\sqtrois\,a_4\sAs_ay_a}$,
$b_2=-\frac{4{a_7}^3{b_4}^2\sAa_a}{\sqtrois\sAs_ay_a}$, $b_3=-\frac{2{a_7}^2
{b_a}^2\sAa_a}{\sqtrois\,a_4y_a},b_5=\frac{b_4a_5}{a_4}$, $b_7=-\frac{b_4a_7}
{a_4}$, $a_8=\frac12-\frac{2{a_7}^2b_4\sAa_a}{y_a}$ and $b_8=\frac12+
\frac{3\sAs_ay_a}{8a_4{a_7}^2\sAa_a}$.

If a third condition $a_7=b_7$ is added, the system follows $b_i={-}a_i$
$\forall i=4,5$ and $b_i=a_i$ $\forall i=6,7$. Its basic equation reads $(x_3)$
$\Aa_a (3-\Aa_a)=3{a_7}^2$ with $\widehat{x_3}=\frac1{4{a_4}^2{a_7}^4}
\widehat{x_2}\rfloor_{b_4\to a_4}= \frac1{4{a_7}^4\sAs_a}\widehat{7_p}
\rfloor^{a_6\to0,b_6\to0,b_7\to a_7}_{b_4\to-a_4,b_5\to-a_5}$. There are 3
degrees of freedom $(a_4,a_5,a_7)$ and the parametrization of other ones reads
$a_1=\frac{2a_5a_7\sAa_a}{\sqtrois\sAs_a}$, $a_2=-\frac{a_4}{a_5}a_1$,
$a_3=\frac{\sAa_a}\sqtrois$, $b_i=a_i$ $\forall i=1..3$, $a_8=
-\frac14+\frac{3{a_7}^2}{4\sAa_a}$ and $b_8=a_8$.

These examples demonstrate the sophistication of algebraic manipulations.
Associativity of the composition of conditions is valid to deduce
basic equations (although a surprising additional factor
$\frac{{a_4}^3}{{b_4}^3}$ emerges depending on the way the last equation is
constructed). But the parametrization is completely different for each case and
associativity cannot be used to deduce it, because the parametrization of the
reduced systems differs from that of the complete one (when projecting this
one following the same conditions). The only way to obtain the correct
parametrization is to use rules \ref{RR}:
\renewcommand{\theequation}{$\mathcal{R}$}
\begin{equation}
\label{RR}
\!\!\!\!\!\!\!\!\!\!\!\!
\forall i=1,2\hbox{ and }\forall\alpha=a,b\qquad
\alpha_i=(-1)^i\frac{\sqrt3\,\ei_\alpha}{2\alpha_8-1}\hbox{ and }
\alpha_3=\frac{\sqrt3\At_\alpha}{2(1-2\alpha_8)},
\end{equation}
which are always valid, except for the two first atypical cases
giving $s_2$; they are also only partially valid for the third atypical case. In
addition to \ref{RR} rules, one must substitute $a_8$ and $b_8$, using
specific rules according to each case.

Eventually, one must be aware that a basic equation can be obtained with two
different parametrizations, defining two separate cases. This occurs for
atypical cases but also in general.


\begin{thebibliography}{9}
\bibitem{prime} Some real factors have no real roots and can be discarded. Other
factors correspond to atypical cases, which are treated apart. We have
eventually dealt with all general prime factors.
\bibitem{F123} One observes that some projections of $s_6$ have singular points
but not $\mathcal{S}$, so $s_6$ should be a flag variety. Since $s_6$ is
constructed as the quotient of $SU(3)$ by relations (\ref{ain}) and (\ref{out}),
it may be isomorphic to the complete flag variety $F_{12}(\C^3)$. One needs to
prove that these relations give the folding rules of a torus $T^2$, but it
exceeds the aims and possibilities of this article.
\bibitem{distance} We define the distance of any point $P$ to $s_6$ as the
absolute value of $\widehat{7_a}$ calculated at $P$: it is indeed a distance,
since it vanishes exactly when $P$ belongs to $s_6$.
\bibitem{close} It would represent a tremendous analysis to ensure the
nontriviality of this hole. Nevertheless, we consider the present characteristic
of $P_t$ as a quasi-determination of its nontriviality.
\bibitem{deter} We have been able to find two families of equations defining
such universal surfaces. In the first one, projections proved too poor for
analysis. In the second one, we have confronted numerical precision
difficulties, which we have not solved yet.
\bibitem{validite} Note that the validity of winding numbers, obtained in
$\widetilde{\mathcal{S}_1}$, holds only thanks to that in $S_1$ and
$\mathcal{S}$ respectively.
\bibitem{detail} Note that (7$_b$) is symmetrical to (\ref{eq7}), so that any
system which can be mapped on (7$_b$) can also be mapped on (\ref{eq7}) using
$a\leftrightarrow b$; therefore (7$_b$) can be omitted.
\end{thebibliography}
\end{document}